\documentclass[11pt]{article}

\usepackage{mathtools}
\usepackage[T1]{fontenc}
\usepackage{float}
\usepackage{subcaption}
\usepackage[textsize=footnotesize]{todonotes}
\usepackage[prefix=sol-]{xcolor-solarized}
\usepackage[algo2e,boxed]{algorithm2e}
\usepackage{graphicx}%
\usepackage{multirow}%
\usepackage{amsmath,amssymb,amsfonts}%
\usepackage{amsthm}%
\usepackage{mathrsfs}%
\usepackage[title]{appendix}%
\usepackage{xcolor}%
\usepackage{textcomp}%
\usepackage{manyfoot}%
\usepackage{booktabs}%
\usepackage{algorithm}%
\usepackage{algorithmicx}%
\usepackage{algpseudocode}%
\usepackage{listings}%
\usepackage{mod}
\usepackage{subfiles} 

\newcommand{\supp}{\operatorname{supp}}
\DeclareMathOperator{\Aut}{Aut}

\lstdefinelanguage{Julia}{
    keywords={function, end, for, if, else, elseif, while, begin, return, struct, mutable, true, false, abstract, primitive, try, catch, finally, in, isa},
    sensitive=true,
    morecomment=[l]\#,
    morestring=[b]"
}

\newcommand\dpoDerivationRaw[6]{%
    \begin{tikzpicture}[gCat, every node/.style={inner sep=1pt, font=\scriptsize}, vCat/.append style={inner sep=.3333em}, node distance=1em and 1em]
       \node[vCat, label=below:$D$]                (D) {#5};
       \node[vCat, label=below:$G$, left=of D]     (G) {#4};
       \node[vCat, label=below:$H$, right=of D]    (H) {#6};
       \draw[eMorphism] (D) to (G);
       \draw[eMorphism] (D) to (H);
       \node[vCat, label=above:$K$, above=of D]    (K) {#2};
       \node[vCat, label=above:$L$, at=(K-|G)]     (L) {#1};
       \node[vCat, label=above:$R$, at=(K-|H)]     (R) {#3};
       \draw[eMorphism] (K) to (L);
       \draw[eMorphism] (K) to (R);

       \draw[eMorphism] (L) to (G);
       \draw[eMorphism] (K) to (D);
       \draw[eMorphism] (R) to (H);
   \end{tikzpicture}%
}

\newcommand\dpoDerivation[7][]{%
	\dpoDerivationRaw%
		{\includegraphics[#1]{#2}}	{\includegraphics[#1]{#3}}	{\includegraphics[#1]{#4}}%
		{\includegraphics[#1]{#5}}	{\includegraphics[#1]{#6}}	{\includegraphics[#1]{#7}}%
}

\newcommand\ruleScale{0.5}
\newcommand\preRuleContent{\scriptsize}

\renewcommand\dpoRuleRaw[3]{%
    \begin{tikzpicture}[gCat, baseline={([yshift=-0.5ex]K.center)}]
        \node[vCat, label=above:$L$] (L) {#1};
        \node[vCat, label=above:$K$, right=of L] (K) {#2};
        \node[vCat, label=above:$R$, right=of K] (R) {#3};
        \draw[eMorphism] (K) to (L);
        \draw[eMorphism] (K) to (R);
    \end{tikzpicture}%
}
\renewcommand\dpoRule[4][]{%
    \dpoRuleRaw{\includegraphics[#1]{#2}}{\includegraphics[#1]{#3}}{\includegraphics[#1]{#4}}
}

\title{Rule-Based Gillespie Simulation of Chemical Systems}

\newcommand\abstractText{
  The MØD computational framework implements rule-based generative
    chemistries as explicit transformations of graphs representing chemical
    structural formulae. Here, we expand MØD by a stochastic simulation
    module that simulates the time evolution of species concentrations
    using Gillespie's well-known stochastic simulation algorithm
    (SSA). This module distinguishes itself among competing implementations
    of rule-based stochastic simulation engines by its atomistic
    explicitness, its flexible network expansion mechanism, and its
    functionality for defining custom reaction rate functions. The last
    point enables direct sampling from actual reactions instead of rules.
    We present methodology and implementation details followed by examples
    which demonstrate the capabilities of the stochastic simulation engine.
    }

\newcommand\ourKeywords{Gillespie simulation, rule-based modeling, on-the-fly network generation}

\usepackage{authblk}
\usepackage{hyperref}
\usepackage[numbers, sort&compress]{natbib}
\date{\today}

\author[1,2]{Erika M. Herrera Machado}
\author[1]{Jakob L. Andersen}
\author[1]{Rolf Fagerberg}
\author[3]{Christoph Flamm}
\author[3,1]{Daniel Merkle}
\author[4--9]{Peter F. Stadler}

\newcommand\email[1]{\texttt{#1}}

\affil[1]{Department of Mathematics and Computer Science,
    University of Southern Denmark, Odense, Denmark
    \email{\{jlandersen,rolf,machado\}@imada.sdu.dk}}
\affil[2]{Faculty of Mathematics and Computer Science, Friedrich Schiller University Jena, Jena, Germany}
\affil[3]{Faculty of Technology, Bielefeld University, Bielefeld, Germany
    \email{daniel.merkle@uni-bielefeld.de}}
\affil[4]{Department of Theoretical Chemistry, University of Vienna, Wien, Austria
    \email{xtof@tbi.univie.ac.at}}

\affil[5]{Bioinformatics Group, Department of Computer Science \&
    Interdisciplinary Center for Bioinformatics \& School for Embedded and
    Composite Artificial Intelligence (SECAI),
    Leipzig University, Leipzig, Germany
    \email{studla@bioinf.uni-leipzig.de}}
\affil[6]{Max Planck Institute for Mathematics in the Sciences,
    Leipzig, Germany}
\affil[7]{Facultad de Ciencias, Universidad National de Colombia,
    Bogot{\'a}, Colombia}
\affil[8]{Center for non-coding RNA in Technology and Health,
    University of Copenhagen, Frederiksberg, Denmark}
\affil[9]{Santa Fe Institute, Santa Fe, NM, USA}

\begin{document}
\maketitle	

\begin{abstract}
\abstractText

\vspace{1em}
\end{abstract}


\section{Introduction}

Mathematical and computational models of chemical and biochemical reaction
networks are indispensable for understanding the behavior of such systems.
While the classical ODE-based approach is suitable for many cases, it may
not accurately represent the true time evolution of a system where
discreteness and stochasticity are important
\cite{bartholomay1958stochastic,gillespie1976general}. This is in
particular the case for systems with very small particle counts, such as
proteins in single cells, which are poorly modeled by real-valued
concentrations \cite{gillespie2007stochastic,mcquarrie1967stochastic}.

Stochastic modeling of chemical kinetics has been explored since the
1940s \cite{kramers1940brownian,montroll1957application,delbruck1940statistical}.
Given a set of reactions $\{R_1, \dots, R_M\}$, stochastic chemical
kinetics is determined by the propensity function of each reaction, which
can be interpreted as a (non-normalized) probability that reaction $R_i$
will occur in the system in the next time interval $[t, t+dt)$. The
reaction probabilities in turn determine the Chemical Master Equation
(CME), a continuous-time Markov process whose state space comprises the
particle counts of all chemical species. Although the CME can be solved
analytically in principle, this is usually impracticable. In particular for
highly nonlinear chemical reactions, even numerical solutions can be
unfeasible.  This motivates the use of stochastic simulation algorithms to
generate sample trajectories of the system's discrete state over time.

Gillespie's stochastic simulation algorithm (SSA)
\cite{gillespie1976general, gillespie1977exact, gillespie2007stochastic}
has become a standard way to simulate chemical and biochemical systems and
serves as the basis for various stochastic simulation tools that generate
trajectories of species concentrations.

In its original version, SSA assumes that the whole set of reactions and
all possible molecular species are explicitly enumerated in advance, which
can be unfeasible in many cases.  For instance, in the field of biochemical
reaction networks, many proteins have multiple sites where chemical
processes can alter them.  A simple heterodimer of two different proteins,
each with eight modifiable sites, for instance, would result in 256
different states, requiring more than 65,000
equations \cite{danos2007rule}.

SSA can be adapted to handle essentially open-ended systems in cases
where it is feasible to, at each time step, enumerate all possible reaction
channels and compute their associated propensities. This setting has been
used frequently in models of RNA evolution with replication rates dependent
on the secondary structures of the parent RNA and mutations introduced
uniformly \cite{Fontana:87,Fontana:89}. A similar application is the
simulation framework ALF for genome evolution \cite{Dalquen:12}.  In the
latter, genomes in the population also interact via lateral gene transfer,
i.e., an offspring of a given parent may also include a piece of DNA copied
from another member of the population.

Rule-based modeling addresses the problem of advance enumeration of the
system in a more general manner by representing species as agents and
interactions as rules that describe how a local pattern should be
transformed \cite{chylek2014rule,chylek2015modeling}.  Since a single rule
can represent a class of reactions, this avoids the need to enumerate all
possible reactions between all possible species
\cite{danos2007rule}. Iteratively applying the rules according to their
propensities can automatically construct the reaction network as part of
the simulation, thus facilitating open-ended systems.

Several specialized simulation frameworks have been developed to
support the rule-based modeling approach.
%
The most popular are those that can
execute models written in the BioNetGen Language (BNGL) or in Kappa.  The
simulation engines NFsim~\cite{sneddon2011efficient} for BioNetGen and
KaSim \cite{boutillier2018kappa} for Kappa are essentially rule-based
versions of Gillespie's SSA. The common theme of these approaches is their
focus on abstract agents characterized by sites that carry state
information and encode distinct interaction capabilities. Through their
sites, agents connect into \emph{site graphs} serving as the (molecular)
entities to which rules are applied. Rules correspondingly specify
transformations for patterns of sites.  Each simulation step comprises
three parts: (i)~calculation of the propensity for each rule based on the
current molecular state; (ii)~sampling of the rule to be applied and the
time step to the next event; (iii)~application of the selected rule and
update of the population of agents
\cite{harris2016bionetgen,boutillier2018kappa}. This scheme avoids the
explicit construction of the reaction network since the decision on the
rule application to be executed can be made directly from the propensities
of all applicable rules.

Using the same principles several alternative
simulation engines have been developed. RuleMonkey
\cite{colvin2010rulemonkey}, like NFsim, is a stochastic simulator for
models written in BNGL.  PISKaS \cite{perez2018stochastic} is a multiscale
simulation tool able to perform stochastic simulations on distributed
memory computing architectures.  It expands the Kappa language by allowing
the explicit declaration of interconnected compartments to simulate
heterogeneous environments and different types of transport between
compartments.  SimSG~\cite{ehmes2019simsg} is a general-purpose tool for
performing rule-based simulations using stochastic graph transformations on
site graphs.  It uses a priority queue to schedule the execution of
rule-match pairs. SYNTAX \cite{cohen1994syntax} is a rule-based stochastic
simulator for metabolic pathways, operating on rules describing the
transfer of carbon atoms from reactants to products.

Site graphs provide a level of abstraction that is well-suited to describe
interactions between macromolecular entities such as DNA strands or
proteins. Models of small molecule chemistry, however, require a more
fine-grained level of resolution in which agents would describe atoms.
Such a level is necessary if one wants to model the fact that chemical
reactions reshuffle the atoms between molecules, for instance in
applications like the design and analysis of stable isotope labeling
experiments where the tracing of individual atoms is important. Very little
is gained by conceptualizing atoms as agents and molecules a site
graphs. Rule-based stochastic simulations that operate \emph{directly} on
molecular graphs would be much more natural for such applications.

We therefore describe here the implementation of a stochastic
simulation engine inside the MØD software framework for
cheminformatics~\cite{andersen2016software}. This engine combines
MØD's rigorous modeling approach for chemistry based on molecular graphs and graph transformation rules with the established stochastic simulation
techniques of SSA. Atomistic explicitness, the ability for sampling
the specific reaction to occur next in the system instead of sampling
a rule, and a rich set of features from which to calculate reaction propensities, are the main strengths and key novelty offered by our simulation engine.

This paper is structured as follows: In Section~\ref{sec: mod}, we
introduce the MØD software. In Section~\ref{sec: gill}, we review
Gillespie's stochastic simulation algorithm (SSA) and how reactions can be
generated as needed. In Section~\ref{sec:features}, we
describe some of the features of
our stochastic simulation engine, while in
Section~\ref{sec: impl}, we provide the details of the implementation. In
Section~\ref{sec: results}, we illustrate the application of the engine
through various examples. Finally, we conclude in Section~\ref{sec:
  conclusion}.

  
\section{The MØD Software for Cheminformatics} \label{sec: mod}

In chemistry, molecules can be represented as undirected graphs where
vertices represent atoms and edges represent bonds
\cite{sylvester1878application,sylvester1878application2}.  Labels associated with each vertex
encode element type and charge, while labels on edges encode the
specific bond type.  In this representation, a chemical reaction can
be modeled as a graph transformation rule
\cite{benko2003graph,rossello2004analysis}. A graph transformation
rule defines how a graph can be modified into a new one. A set of
transformation rules together with a set of initial graphs is
referred to as a graph grammar \cite{rozenberg1997handbook}. Used in
the context of chemistry, rules are applied to a set of initial
molecular graphs to yield new molecular graphs through the creation
and destruction of bonds.  By the Law of Conservation of Mass,
vertices (i.e., atoms) can never be destroyed or created.

MØD \cite{andersen2016software} is a software package that implements a
chemically inspired graph transformation system based on the Double
Pushout (DPO) formalism \cite{rozenberg1997handbook} for graph
transformation.  See Fig.~\ref{fig: dpo} for an example of a DPO
transformation rule and its application to a molecular graph.
In Fig.~\ref{fig: dpo}, the appearance of the left-hand side~$L$ of the rule inside the molecular graph~$G$ is called a \emph{match} of the rule.
A rule can be applied to any graph~$G$ for which such a match exists.
In particular, a rule is able to model a class of reactions with the same overall molecular change but with different educt molecules.
The specificity of a rule can be controlled by varying the graph~$K$ which determines the part of the match that is not changed during the rule application.

\begin{figure}
  \centering
  \def\pt{figures/dpo/out/002_r_4_10300000_0_der}
  \def\pb{figures/dpo/out/004_r_6_10300000_0_der}
  \scriptsize
  \dpoDerivation[scale=0.5]{\pt L}{\pt K}{\pt R}{\pb G}{\pb D}{\pb H}
   
  \caption{A Double-Pushout transformation representing a reverse aldol
    reaction.  The rule is the top so-called span $r = \left( L \leftarrow
    K \rightarrow R \right)$, which is applied to an input molecular
    graph~$G$ inside which $L$ appears as a subgraph. The transformation to
    the output~$H$ proceeds by first deleting the difference between $L$
    and $K$ from $G$ to obtain $D$, and then adding the difference between
    $R$ and $K$ to $D$ to obtain the product graph~$H$.  In general, the
    graphs $G$ and $H$ may consist of multiple connected components, each
    being a molecular graph representing educt/product mocules.  In
    the case above, $H$ has two connected components, $H_1$ and $H_2$.
    The rule application thus generates a reaction
    $G\longrightarrow H_1 + H_2$.}
  \label{fig: dpo}
\end{figure}

In contrast to the Single Pushout (SPO) approach, used for instance in
Kappa, the DPO approach has no side effects during rule application,
making the transformations invertible. This is more suitable for
low-level chemistry, where individual atoms are tracked and chemical
reactions are in principle reversible. For detailed information on how the
graph transformation engine is implemented in MØD, we refer to
\cite{andersen2016software}.

Although MØD implements DPO graph
transformation in full generality, it provides many features
specifically designed for handling chemical data. It is also
particularly effective for exploring and generating potential reaction
networks by repeatedly applying rules, thereby evolving a set of
molecular species. Additionally, it includes algorithms for composing
transformation rules, which can be used, for example, to abstract
reaction mechanisms or entire trajectories into overall rules
\cite{andersen201450}. In addition to generic graph data, the software
can load graphs from SMILES strings and other chemical formats, and
allows for visualizations of graphs, rules, and DPO diagrams in a
manner similar to how molecules are usually depicted in
chemistry. MØD's general capability to also model abstract (that is, not
necessarily chemical) graphs can itself be useful for applications in
chemistry, e.g., when modeling chemistry where not all details are
necessary and parts of molecules can be merged into vertices with
non-chemical labels.

\section{Gillespie's Stochastic Simulation Algorithm and
 Reaction Generation on Demand} \label{sec: gill}

For completeness, we briefly recapitulate Gillespie's stochastic
simulation algorithm (SSA) \cite{gillespie1976general}.
Gillespie describes the SSA as ``a Monte Carlo procedure
for numerically generating time trajectories of the molecular
populations in exact accordance with the Chemical Master Equation''
\cite{gillespie2007stochastic}. The simulation generates successive
states of the system, where a state is a vector of non-negative
integers specifying the molecular population, i.e., the number of each
molecular species.  Given such a state $x$, each reaction $R_i$ has
propensity $a_i(x)$, while the sum of all propensities is denoted
$a_{\mathit{tot}}(x)$.  Each iteration of the simulation samples the next
reaction to carry out according to the propensities and samples
the step for advancing time from an exponential distribution with
mean $1/a_{\mathit{tot}}(x)$.  Specifically, two numbers $r_1$ and $r_2$ are
drawn from a uniform distribution over $[0,1]$, and the index of the
next reaction is then calculated as the smallest integer $j$ satisfying
$\sum_{i=1}^j a_{i}(x) > r_1 \cdot a_{\mathit{tot}}(x)$ and the time step to the
next reaction is calculated as $\tau = (-\ln r_2)/a_{\mathit{tot}}(x)$.
Each reaction $R_j$ corresponds to a state change vector
$\nu_j$, and the next state is thus $x + \nu_j$.
Usually, the propensity~$a_i(x)$ in state~$x$ of reaction~$R_i$ is
calculated according to the law of mass action.

The original Gillespie SSA is formulated assuming that the reaction network
is fixed \textit{a priori}. Rule-based approaches allows for a more dynamic
view of the network: At any given instant, there is a finite population of
molecules. This in turn makes some specific subset of the rules applicable,
where each potential rule application (a rule may have more than one)
corresponds to a reaction. A natural approach is therefore in each
simulation step to generate the currently possible reactions \emph{on
demand}, based on the rules and the current population of molecules. This
methodology can significantly speed up the simulation in cases where the a
priori network is large but the simulation only touches a small part of it
at any given time. In some cases, such as polymer systems with an inflow of
the base molecule(s), the \textit{a priori} reaction network may even be
infinite, making a rule-based methodology with reactions created on demand
a strict necessity. We therefore arrive at Algorithm~\ref{alg:generic-ssa},
where the extra step responsible for generating reactions on demand is
marked by color.

\SetAlCapSkip{2ex} 
\begin{algorithm2e}
  \caption{Gillespie's stochastic simulation algorithm (SSA). Here,
    $x$ is the state of the current step, and $t$ is the position on
    the simulation timeline of the current step. The
    threshold~$t_{max}$ controls the length of the simulation
    timeline. The colored line is the generation of
    reactions on demand. In Section~\ref{sec: impl}, we give details of
    its implementation in MØD.}
  \newcommand\AdHoc[1]{{\color{sol-magenta}#1}}
\DontPrintSemicolon
\KwIn{$x_0$, $t_0$, and $t_{max}$}
$x\leftarrow x_0$;\,
$t\leftarrow t_0$\;
\While{$t_k \leq t_{max}$} {
  \AdHoc{Generate the reactions possible in~$x$}\;
  Calculate propensities $a_i(x)$ for each reaction $R_i$ and calculate
  $a_{\mathit{tot}}(x)$\;
  Generate time step $\tau$ and reaction index $j$\;
  Compute $\nu_j$\;
  $t \leftarrow t + \tau$;\, 
  $x \leftarrow x + \nu_j$\;
  Add $(x, t, j)$ to simulation trace\;
}
\label{alg:generic-ssa}
\end{algorithm2e}%

The extreme version of this approach, sometimes termed \emph{network free}, is to  calculate the set of reactions anew in each simulation step.
In Section~\ref{sec: impl}, we describe how
we use caching schemes to allow a gradual interpolation between the classic
\textit{a priori} view and the network free view. This interpolation
results in a trade-off between speed and memory usage which can be tuned to
the characteristics of the chemistry simulated.

We note that in both BioNetGen~\cite{faeder2009rule} and
KaSim~\cite{boutillier2018kappa}, the rules are sampled according to their
numbers of matches in the current state~\cite{suderman2019generalizing}.
Rate constants are thus not defined per reaction, but are associated to rules,
and all reactions generated by the same rule therefore have the same
rate constant. In Section~\ref{sec:features}, we describe how our simulation engine based on MØD is significantly more flexible in this regard.



\section{Some Features of the Simulation Engine Interface}
\label{sec:features}

The approach we take for our stochastic engine based on MØD is that we
explicitly generate reactions in a rule-based manner and give the user the
option to calculate a rate constant for each reaction through a custom
rate-constant function. Thus, the user retains full control over the
reaction kinetics, which in particular can depend not only on the reactant
molecules but also on the specific match of the reaction rule inside the
reactant molecules. To this end, the rate-constant function has access to
the graph structure of each molecule participating in a given reaction, to
the rule which was used to generate the reaction, and to where in the
molecules the rule matched. The rate-constant calculation can be as
detailed as desired, from simply returning a fixed rate constant based on
the rule to invoking quantum chemical tools.  In programming terms, the
rate-constant function is implemented as a callback. Additional callbacks
are available that provide the user with access to further information such
as the current state and iteration number.  Below, we exemplify by two
particular use cases the possibilities enabled by this detailed interface.

\subsection{Handling Symmetry and Effective Rule Activity}

Each reaction is generated by the application of a rule $r = (L\leftarrow K\rightarrow R)$ with a match $m$ of $L$ inside an input graph $G$
to create an output graph $H$, as shown in Fig.~\ref{fig: dpo}.
Formally, such a match is a graph monomorphism $m\colon L \rightarrow G$.
\emph{Symmetries} of $L$ and $G$ are graph automorphisms, i.e., isomorphisms $L\rightarrow L$ and $G\rightarrow G$,
while symmetries of $r$ are triples $\langle \alpha_L, \alpha_K, \alpha_R\rangle$ of automorphisms of the three graphs of the rule
such that they commute with the rule span itself, i.e., the following diagram commutes.
\begin{center}
\begin{tikzpicture}[mor/.style={->, >=stealth}]
\matrix[matrix of math nodes, row sep=2em, column sep=2em]{
	|(L1)| L 	& |(K1)| K 	& |(R1)| R 	\\
	|(L2)| L 	& |(K2)| K 	& |(R2)| R 	\\
};
\draw[mor] (K1) to (L1);
\draw[mor] (K1) to (R1);
\draw[mor] (K2) to (L2);
\draw[mor] (K2) to (R2);
\draw[mor] (L1) to node[left] {$\alpha_L$} (L2);
\draw[mor] (K1) to node[left] {$\alpha_K$} (K2);
\draw[mor] (R1) to node[left] {$\alpha_R$} (R2);
\end{tikzpicture}
\end{center}
Different matches may be considered equivalent when symmetries are taken into account,
and one thus has a choice of whether to treat all matches as
separate events or whether to treat them collectively according to some or
all of the symmetries, effectively ``collapsing'' them in a symmetry-aware
way. In Kappa for instance, three stances are distinguished
\cite[Sec.~3.3]{kappaManual}:
\begin{enumerate}
\item Consider matches that are isomorphic according to the symmetries of
  $L$ as a single event.  Formally, $m_1, m_2\colon L\rightarrow G$ are
  equivalent if for any $\alpha\in \Aut(L)$ we have $m_1\circ\, \alpha =
  m_2$. In Kappa, this is referred to as the ``non-deterministic''
  interpretation of the rule action.
\item Consider matches that are isomorphic according to the symmetries
  of the whole rule $r$ as a single event.  Formally,
  $m_1, m_2\colon L\rightarrow G$ are equivalent if for any
  $\langle \alpha_L, \alpha_K, \alpha_R\rangle \in \Aut(r)$ we have
  $m_1\circ\, \alpha_L = m_2$. In Kappa, this is referred to as the
  ``deterministic'' interpretation of the rule action.
\item Do not consider matches to be equivalent, but see them as distinct
  events.  In Kappa, this is referred to as the ``null'' interpretation.
\end{enumerate}
Note that Kappa does not take symmetries of $G$ into account,
hence two matches $m_1, m_2\colon L\rightarrow G$ will generate two events even when there is a symmetry $\alpha\in \Aut(G)$
such that $\alpha\circ m_1 = m_2$. Thus, further stances than the three listed above are possible.

In MØD, the starting point is that a reaction event is defined only based
on $G$ and $H$, independently of the rule(s) used to create it and how they
could match, similarly to how reactions are defined in the standard SSA.
The user can then use the reaction rate constant callback to implement
their preferred stance, using e.g.\ the functionality of MØD to enumerate matches and
symmetries.

\subsection{Atom Tracing and Match-Specific Events}
In addition, having full access to each possible match for a given reaction event can be crucial for atom
tracing in simulations, such as those motivated by stable-isotope labeling
experiments.  In these settings, the location of a particular atom (e.g., a
labelled carbon) might follow different trajectories depending on which
matches one considers to have occurred.  Because MØD can
distinguish between all possible ways of applying a rule, a callback can
incorporate any external logic needed to track particular atoms or to
weigh matches differently for the propensity calculation.
As the result of the simulation is a trace of
reactions with full access to molecular structures, the isotope
tracing can also be deferred to a post-processing step.  These capabilities
allow the simulation
engine to track individual atoms throughout the reaction network,
providing a powerful tool for research that relies on identifying the
fate of specific atoms in chemical or biochemical pathways.

\section{Implementation} \label{sec: impl}

In a rule-based system with a set of rules~$\mathcal{R}$ and a state~$x$,
each match of a rule $r \in \mathcal{R}$ on a multiset of molecules present
in~$x$ defines a possible reaction. A rule may have several matches, even
on the same multiset. To perform a simulation step in
Algorithm~\ref{alg:generic-ssa}, we must determine the set of possible
reactions in~$x$.

Testing for matches between a rule and a multiset of molecules (and in case
of a match also applying the rule to generate the associated reaction) is a
time-consuming operation. Performing this for all rules and all multisets
in every step of the simulation can slow down the algorithm significantly.
Our implementation therefore employs several caching mechanisms.

Denote by $\supp(x)$ the subset of molecules of state~$x$ with non-zero
count. Clearly, a reaction is only possible if all molecules involved in
its underlying rule match (i.e., all molecules having non-zero
stoichiometric coefficients for the reaction) are in $\supp(x)$. The
principle of our caching schemes is that they maintain a superset of all
reactions fulfilling this criterion. Each reaction in the cache stores its
stoichiometric coefficients, so calculating its propensity from the
specified rate constant and the state vector~$x$ of molecular counts is
fast and does not require any further tests for matching.

The goal of our first caching scheme is to avoid the test between all rules
and all multisets in each iteration by only looking at potential new
reactions compared to the last step.  This is done by leveraging the
strategy framework described in \cite{andersen2014generic}.  The core of
the framework is controlling the application of graph transformation rules
with the help of two sets of molecular graphs: the universe $\mathcal{U}$
which comprises all molecular graphs that are available for transformation
(i.e., $\mathcal{U} = \supp(x)$), and the distinguished subset
$\mathcal{S}\subseteq\mathcal{U}$ which identifies the part of the universe
that must participate in every new reaction to be generated.  Each reaction
is generated by the application of a rule $r$ to a graph $G$ of reactants.
The graph $G$ is constructed as the disjoint union of a multiset of
molecular graphs $\mathcal{G} = \{g_1, g_2, \dots, g_k\}$.  Given a
universe-subset pair $(\mathcal{U}, \mathcal{S})$ the framework will for
each rule $r \in \mathcal{R}$ generate all such multisets where
$\mathcal{G} \subseteq \mathcal{U}$, $\mathcal{G}\cap \mathcal{S} \neq
\emptyset$, and $k$ is at most the number of connected components of the
left-hand side~$L$ of the rule $r = \left( L \leftarrow K \rightarrow R
\right)$.  All matches and corresponding transformations with the rule are
then computed to generate reactions.

This distinction between $\mathcal{U}$ and $\mathcal{S}$ provides a
structured way to store newly derived molecules and thereby target rule
application to avoid recomputation of reactions.  In the initial state,
$x_0$, apply rules using the universe-subset pair $\mathcal{U}_0 =
\mathcal{S}_0 = \supp(x_0)$, and add all reactions found to the cache.  In
step $i$ with state $x_i$, apply rules with $\mathcal{U}_i = \supp(x_i)$
and $\mathcal{S}_i = \supp(x_i)\backslash \supp(x_{i-1})$, and add all
reactions found to the cache.  That is, let the subset $\mathcal{S}_i$ be
all the molecular graphs that have non-zero count in this iteration but
count zero in the previous iteration, thereby only calculating reactions
where those new molecular graphs participate.  Note that if $\mathcal{S}$
is empty, then no new reactions are generated.  By induction, in each
iteration all reactions possible in~$x$ are available in the cache.

Our second caching strategy is based on the following observation:
if iteration $i$ starts with state $x_i$ with support $\supp(x_i)$ and a subsequent iteration $i'$ has support $\mathcal{U}_{i'}\subseteq \mathcal{U}_i$,
then all relevant reactions will already by present in the cache.
Our implementation therefore caches each $\supp(x_i)$ that were used for rule application to test this subset property and thereby avoid redundant calculations.

Importantly, these caching mechanisms only serve to speed up the algorithm,
and thus if a user notices that memory usage becomes large (for example, in
complex systems that expand into a large number of intermediate states),
there is a straightforward option to ``restart'' the reaction generation by
clearing the caches.  Any subsequent iterations again generate new
reactions and store them in fresh caches.  This gives users direct control
over balancing performance improvements from reusing reaction generation
against memory constraints that can arise from storing large amounts of
state information.

Under the hood, the core of the strategy system, the actual computation of
graph embeddings, and the reaction data structures, are all implemented in
C++~\cite{andersen2014generic,andersen2016software}.  The morphism
algorithms that search for valid embeddings of the left-hand side of a rule
in a molecular graph are thus executed efficiently.  User-defined
rate-constant callback functions, which may depend on the number of
embeddings or on specific topological properties of the molecules, can be
provided via a Python-based interface, allowing flexible customization.  By
switching to pre-compiled callbacks, performance can be optimized further
in time-critical scenarios.

\section{Illustrative Examples of Application} \label{sec: results}

This section demonstrates the practical capabilities of the M{\O}D
stochastic simulation engine using two chemically grounded case
studies. First, we reproduce the well-characterized anaerobic degradation
of atrazine and use a callback to transition from a closed to an open
system during runtime, highlighting how the engine can alter boundary
conditions in response to internal simulation events. Second, we move to a
generative setting and simulate wet-dry cycling chemistry leading to
depsipeptide formation. In this example, the reaction networks are created
on-the-fly and environmental hydration is controlled dynamically via
callbacks. Together, these two applications illustrate rule-based on-demand
reaction generation, dynamic environmental control, and \textit{post hoc}
analysis from the full stochastic event trace—capabilities that go beyond
classical approaches with enumerated reaction spaces.

\subsection{Degradation of Atrazine}

\begin{figure}[tbp]
  \centering
  \includegraphics[width=\textwidth]{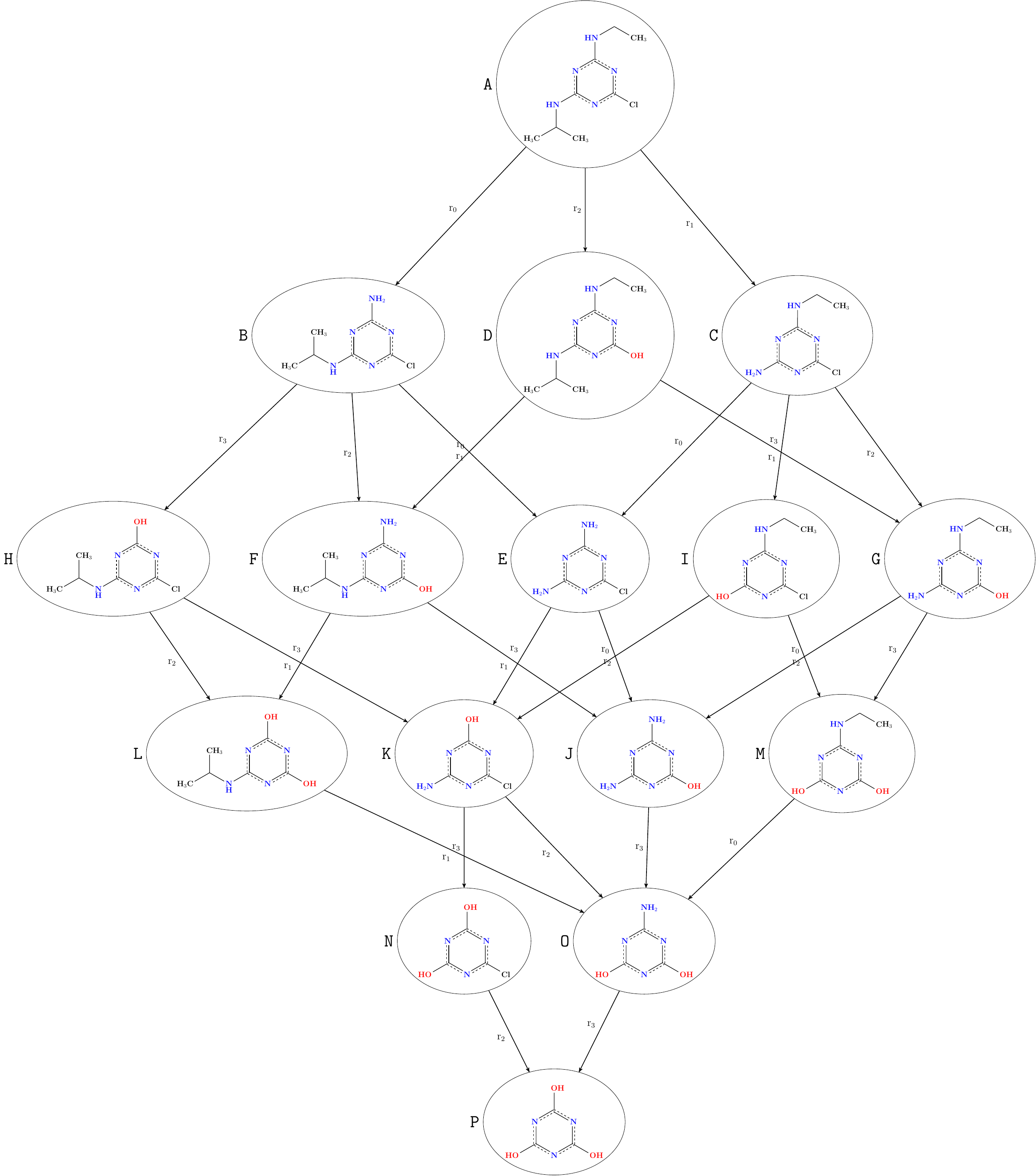}
  \caption{Atrazine degradation network depicting all 
    intermediate products.  The initial compound \texttt{A} is atrazine,
    and \texttt{P} is cyanuric acid.}
  \label{fig:network}
\end{figure}

Our first example illustrates how the stochastic simulation engine in
M{\O}D can be used to reproduce chemically realistic degradation dynamics,
while also exploiting callback mechanisms to alter the simulation
conditions mid–execution.  We consider the well–studied anaerobic
degradation of atrazine, a widely used herbicide of the s–triazine family.
Under anaerobic conditions in soil, atrazine degrades stepwise to cyanuric
acid through repeated hydrolysis and reductive dealkylation
\cite{erickson1989degradation}:
\begin{itemize}
\item[] \textbf{Hydrolysis:} \quad $\mathrm{R{-}X \xrightarrow{H_2O}
  R{-}OH + HX}$ \quad (where $X=\mathrm{Cl}$ or $\mathrm{NH_2}$).
\item[] \textbf{Reductive dealkylation:} \quad $\mathrm{R{-}NH{-}R'
  \xrightarrow{H_2} R{-}NH_2 + HR'}$ \quad (with $R'$ a short ($\mathrm{C_1{-}C_4}$) alkyl residue such as ethyl or
  isopropyl).
\end{itemize}
Repeated application of these two reaction types generates a
network of intermediates leading from atrazine (\texttt{A}) to cyanuric
acid (\texttt{P}), including chloro-, amino-, ethyl-, and
isopropyl-substituted species. The stochastic simulation is not provided
this network in advance: only the initial molecule and the
transformation rules are given, and all intermediates arise dynamically as
rule matches are discovered during the simulation. For illustration, we
assign the labels \texttt{A}--\texttt{P} \textit{post hoc} to the distinct
structures that emerge and depict in Fig.~\ref{fig:network} the resulting
complete network. Note that the same methodology applies equally well to
substantially larger and more combinatorial chemical spaces where explicit
enumeration would no longer be feasible.

To model the reaction kinetics we use the experimentally determined rate
constants reported in \cite{erickson1989degradation}:
$k_{\text{\tiny hydrolysis}} = 5.00\times10^{-9}\,\mathrm{s}^{-1},$
$k_{\text{\tiny de-ethylation}} = 3.32\times10^{-8}\,\mathrm{s}^{-1}$,
and
$k_{\text{\tiny de-isopropylation}} = 2.65\times10^{-8}\,\mathrm{s}^{-1}$.

\begin{figure}
  \centering
  \includegraphics[width=0.7\textwidth]{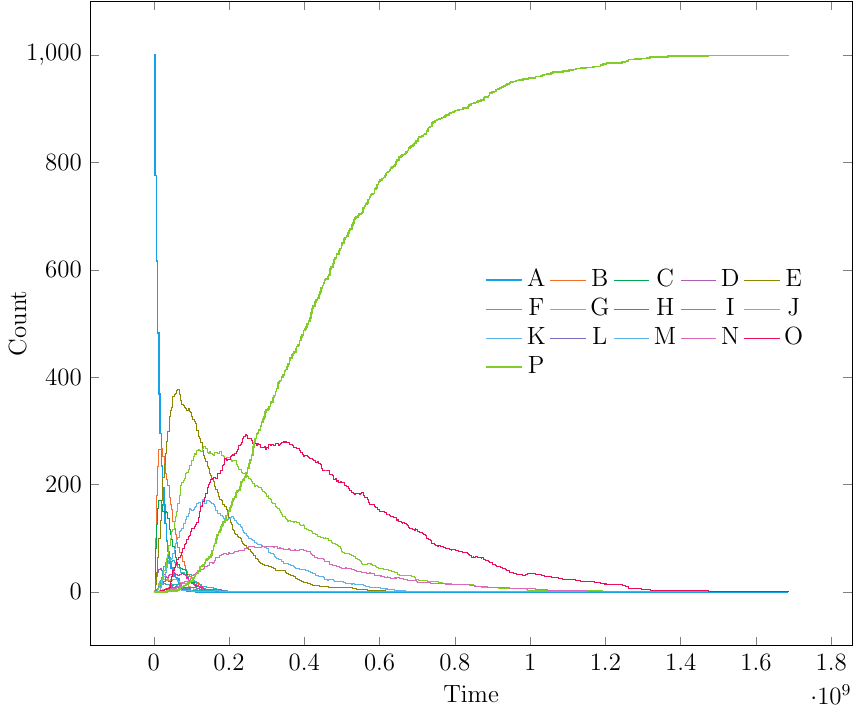}
  \caption{Stochastic simulation of the atrazine degradation example.
   The simulation is in a closed system,
   starting with $\texttt{A}=1000$, $\text{\texttt{B}--\texttt{P}}=0$ and no inflow or outflow.
   Approximately at time $1.67\times10^9$ the system reaches equilibrium with
   $\text{\texttt{A}--\texttt{N}}=0$ and $\texttt{P}=1000$,
   where no further rule applications are possible.
 }
  \label{fig:atrazine-sim-closed}
\end{figure}

\paragraph{Closed–system stochastic simulation}
We first run the degradation in a closed system, i.e., without inflow or
outflow of any species. The simulation starts from a population of atrazine
molecules (\texttt{A}) and applies the four graph–transformation rules in
Fig.~\ref{fig:atrazine_rules} on-the-fly. Reaction events are scheduled by
Gillespie's SSA, with the fixed rate constants stated above provided via the rate-constant callback of the simulation engine. Because no new substrate enters and no product leaves, the trajectory converges to the thermodynamic equilibrium expected for a closed system: all mass accumulates in cyanuric acid (\texttt{P}). At this point no further matches exist, all propensities drop to zero, and the simulation reaches a deadlock.
The course of the simulation is depicted in Fig.~\ref{fig:atrazine-sim-closed}.
\begin{figure}
  \centering
  \input{figures/rulesAtrazine/inc.tex}
  \caption{Graph transformation rules for the degradation of atrazine.
    Each rule replaces a specific substructure in the left–hand side with a
    chemically valid product pattern in the right–hand side, preserving the
    atom labels and mass balance.}
  \label{fig:atrazine_rules}
\end{figure}

For completeness, we also implemented a traditional ODE model of the
closed reaction network using Julia. The ODE formulation and the corresponding
deterministic simulation results are included in Appendix~\ref{appendix:ODE}, where they serve as a baseline demonstrating the fidelity of our stochastic simulation engine.

\paragraph{Callback–controlled transition to an open system}
The stochastic engine in M{\O}D allows users to change the physical conditions of the simulation whenever a specified event occurs. In this example, we use a rate-constant callback that monitors the state of the simulation in order to detect when the deadlock of the closed system is reached (i.e., when no further reaction events are possible). At this point, the callback changes its rate-constant answers, switching the system from closed to open by activating two new pseudo–reactions:
\begin{itemize}
\item Constant \emph{inflow} of atrazine (\texttt{A}) into the system.
\item Constant \emph{outflow} (removal) of cyanuric acid (\texttt{P}).
\end{itemize}
By a pseudo–reaction, we mean a reaction where either the left-hand side (inflow) or the right-hand side (outflow) is empty. Inflow is a zero-order reaction while outflow is a first-order reaction.

In other words, as soon as the system has degraded all initial atrazine into
\texttt{P}, new atrazine begins entering the reactor while the final
product is simultaneously removed.  The result is a qualitatively different
dynamical regime: the former equilibrium of the closed system becomes
unstable, and the system evolves toward a non–equilibrium steady state
(NESS) in which several intermediates coexist with \texttt{P} at steady
concentrations.

\begin{figure}
  \centering
  \includegraphics[width=0.7\textwidth]{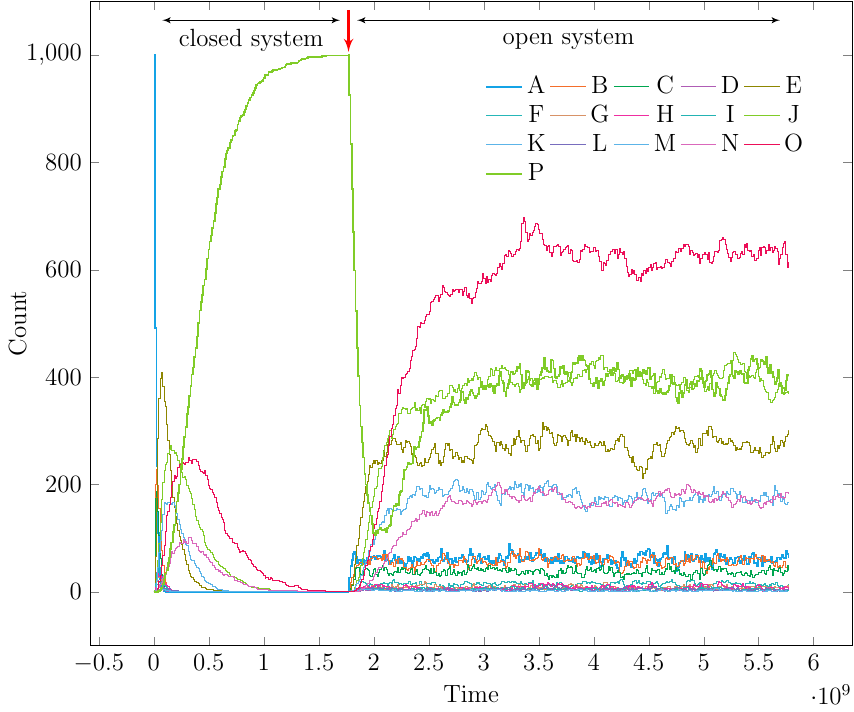}
  \caption{Stochastic simulation of the atrazine degradation example.
   The simulation starts in the closed system, as in Fig.~\ref{fig:atrazine-sim-closed}
   ($\texttt{A}=1000$, $\text{\texttt{B}--\texttt{P}}=0$, no inflow or outflow),
   and as in Fig.~\ref{fig:atrazine-sim-closed} reaches equilibrium at time~$1.67\times10^9$, indicated by the red arrow. This deadlock event of the simulation triggers the rate-constant callback to change its response,
   which opens the system by enabling inflow of compound
    \texttt{A} (thick cyan line) and outflow of \texttt{P} (thick green
    line). In the now open system the conservation relation (i.e., the sum
    of all species counts being constant) is lost. Therefore the former
    equilibrium state becomes unstable and the open system evolves
    to a non–equilibrium steady state (NESS), where \texttt{P} co-exists
    with other species in the network.}
  \label{fig:atrazine-sim}
\end{figure}

Fig.~\ref{fig:atrazine-sim} shows the complete time course.  The
closed–system phase proceeds until approximately $1.67\times10^9$ simulation time units, where the deadlock is detected.  At that moment, the callback
activates inflow and outflow.  After a short transient, the trajectory
settles into a new long–term behavior characteristic of an open system with
continuous substrate replenishment.

This example highlights several key features of the M{\O}D stochastic engine:
it enables rule-based generation of reactions at the atomistic level;
it supports callback-driven interventions in response to simulation events (here, switching from a closed to an open system);
and it allows seamless continuation of the SSA trajectory after changes to kinetic or boundary conditions.
The ability to react
programmatically to internal simulation states is particularly useful when
environmental changes depend on the system’s chemical status (e.g.,
substrate depletion, critical-concentration thresholds, or timed
triggers). Moreover, because all reactions are stored, the
complete event trace can later be analyzed quantitatively, for instance to
compute residence times of intermediates or to assess how inflow/outflow
policies shape the steady-state composition.

\subsection{Stochastic Wet-Dry Cycling Simulation of Depsipeptide Formation}
\label{sec:stochastic-sim-depsi}

As our second example, we provide a flexible, stochastic, on-the-fly
generative simulation of wet-dry cycling chemistry. The objective is to
capture, within a single configurable framework, the interplay of ester
formation in concentrating (dry/hot) phases and ester-amide exchange plus
differential hydrolysis in rehydration (wet/cool) phases that has been
observed for mixtures of $\alpha$-hydroxy acids and $\alpha$-amino
acids~\cite{forsythe2015main, forsythe2015si}. Rather than aiming to
replicate specific experiments, the emphasis here is on a modular setup in
which chemistry and environment are cleanly separated so that environmental
schedules and control policies can be varied easily via callbacks in an
open system. Wet-dry cycling is informative because rate hierarchies invert
between phases in a chemically interpretable way: during dry phases,
esterification and aminolysis of esters are favored by concentration, while
during wet phases, esters are selectively hydrolyzed and amides remain
comparatively stable. Iteration therefore couples a short-time,
phase-locked oscillation of instantaneous formation rates (often anti-phase
between water level and amide formation) with a potential long-time trend
in amide content, in particular in open systems with substrate
replenishment \cite{forsythe2015main}.

In our implementation, environmental control is effected entirely through
callbacks that vary the inflow/outflow rate constant of water; all other rate constants remain fixed. In particular, water inflow is governed by a
simple hysteretic controller acting on the input rate constant: when the water count exceeds an upper bound, inflow is disabled; when it falls below a lower
bound, inflow resumes. We keep a constant inflow of 
the amino acids glycine and alanine, and the alpha-hydroxy acids glycolic
acid and lactic acid. Fixed output rate constants provide a constant evaporation/outflow background for selected species, in this case water, amino acids and alpha-hydroxy acids.

This realizes an open system with a threshold-driven
hydration policy while keeping the chemistry itself phase-agnostic.
The reaction network is generated on-the-fly during simulation,
and to avoid unbounded growth
we restrict polymer motifs via a predicate: we set an upper bound for the
number of amide- plus ester-patterns to be allowed in a molecule.
The primary readouts are time series of amide bond and ester bond counts
together with the water count, which enables quantitative analyses such as
cross-correlation with the hydration policy and estimation of phase lags.

\begin{figure}
  \centering
  \preRuleContent
  \dpoRule[scale=\ruleScale]{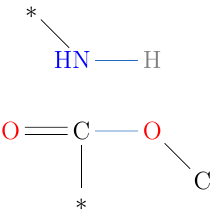}{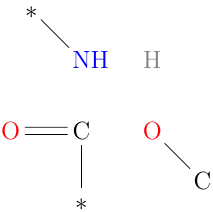}{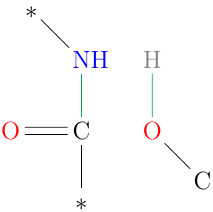}
  \caption{Aminolysis of an ester, modeled as a DPO rule. Stars indicate general wildcards.}
  \label{fig:dpo-aminolysis}
\end{figure}

\paragraph{Rule-based chemistry}
We encode the chemistry with five double-pushout rules over labeled
molecular graphs with explicit hydrogens; the rules are phase-agnostic and
their rate constants remain fixed. In Fig.~\ref{fig:dpo-aminolysis} we
illustrate one of the five rules, aminolysis of an ester, in which an acyl ester
$\mathrm{R{-}C(=O){-}O{-}R'}$ reacts with a primary amine
$\mathrm{H{-}N(H){-}R''}$ to give the amide $\mathrm{R{-}C(=O){-}NH{-}R''}$
and the alcohol $\mathrm{H{-}O{-}R'}$.
The remaining rules model the following reactions:
\begin{itemize}
\item Condensation of a carboxylic acid with an alcohol to form an ester and water.	
\item The reverse reaction, ester hydrolysis by water.	
\item Direct amidation of a carboxylic acid to form an amide and water (included with a low baseline rate constant).	
\item Amide hydrolysis back to acid and amine.	
\end{itemize}
All five rules are detailed in Appendix~\ref{appendix:rules}.
Both inter- and intramolecular reactions can arise depending on the matches of rules in the current simulation state, e.g., allowing lactide-like cyclizations.
Exploratory runs were conducted to find sensible limits not changing the chemical semantics for use in the predicate for avoiding excessive polymer growth.

We use fixed per-rule rate constants for the five rules (esterification
$0.03$, ester hydrolysis $0.01$, aminolysis of esters $0.3$, direct acid
amidation $10^{-5}$, amide hydrolysis as indicated in each run), together
with the above-mentioned
open environment which applies a small constant inflow for monomers
and a hysteretic inflow for water, plus a constant outflow background. These
choices enforce the qualitative ordering that underpins wet-dry behavior:
esterification and especially aminolysis dominate under concentrating
conditions, whereas ester hydrolysis is favored upon rehydration due to
mass-action with abundant water, and amide hydrolysis remains comparatively
slow \cite{forsythe2015main,forsythe2015si}. Initial counts are set to 100
for each monomer species.
These settings match the experimental narrative at a qualitative level while keeping the stochastic network generation tractable.

\paragraph{Illustrative rate-constant variation and outcomes}
%
While the resulting simulation traces would admit many avenues of
quantitative analysis (e.g., cross-correlation between hydration and
amide formation, phase-lag estimation, residence-time and burst
statistics, amplitude and period variability across cycles, and
growth metrics of the evolving reaction graph), our aim here is to
highlight the integration of a rule-based approach with stochastic,
on-the-fly network generation using callbacks and trace
post-processing. We choose to illustrate this integration with a
single controlled change: we vary only the amide-hydrolysis rate
constant while keeping all other reaction and flow parameters at
baseline.

\begin{figure}[t]
  \centering
  \begin{subfigure}[t]{0.32\linewidth}
    \centering
    \includegraphics[width=\linewidth]{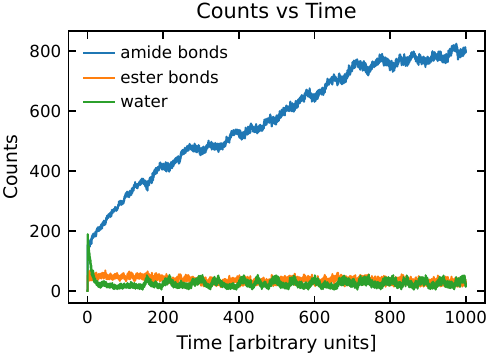}
    \subcaption{Amide-hydrolysis rate constant $0.01$ (accumulation).}
    \label{fig:rate-scan-a}
  \end{subfigure}\hfill
  \begin{subfigure}[t]{0.32\linewidth}
    \centering
    \includegraphics[width=\linewidth]{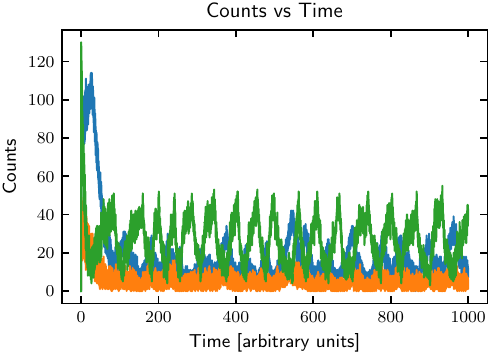}
    \subcaption{Amide-hydrolysis rate constant $0.1$ (oscillation).}
    \label{fig:rate-scan-b}
  \end{subfigure}\hfill
  \begin{subfigure}[t]{0.32\linewidth}
    \centering \includegraphics[width=\linewidth]{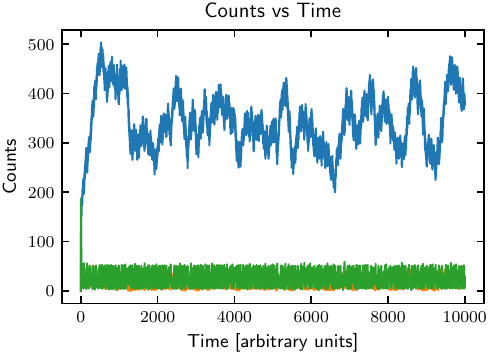}
    \subcaption{Amide-hydrolysis rate constant $0.015$ (transition;
      $t_{\max}=10{,}000$).}
    \label{fig:rate-scan-c}
  \end{subfigure}
  \caption{Open-system wet-dry simulation via callback-controlled hydration. Panels show amide-bond count vs.\ time
    together with water-level traces (not shown in detail here) for three
    choices of the amide-hydrolysis rate constant. Panels (a,b) run to
    $t_{\max}=1000$; panel (c) runs to $t_{\max}=10{,}000$ to reveal
    long-term behavior.}
  \label{fig:rate-scan}
\end{figure}

At a low amide–hydrolysis rate constant of $0.01$
(Fig.~\ref{fig:rate-scan-a}), the amide-bond count
increases more or less monotonically across cycles, with short-time
oscillations still apparent in the instantaneous formation rates. As
expected for an open system with inflow, the reaction network expands and
repeatedly revisits the same motifs, so that multiple copies of identical
compounds accumulate. Ester bonds are progressively drained by aminolysis
during concentrating phases and by hydrolysis during rehydration, whereas
formed amides are only weakly eroded at this rate-constant setting, leading to a net increase in the amide fraction per cycle and a clear long-term drift.

Raising the amide–hydrolysis rate constant to $0.1$ produces a robust anti-phase
relation between water level and the amide-bond count
(Fig.~\ref{fig:rate-scan-b}): amide formation peaks
in low-water intervals and declines upon rehydration, and the series
fluctuates around a nearly stationary long-term mean. The ester pool shows
the complementary behavior. This regime is consistent with wet-dry cycling
reports for mixtures of $\alpha$-hydroxy and $\alpha$-amino acids, where
esters are selectively hydrolyzed in wet phases and amides remain
comparatively stable but can be turned over when hydrolysis is sufficiently
fast \cite{forsythe2015main}.

An intermediate amide-hydrolysis rate constant of $0.015$ yields a transition regime with mixed long- and short-time dynamics
(Fig.~\ref{fig:rate-scan-c}). We extend this run to
$t_{\max}=10{,}000$ (versus $1000$ in
Fig.~\ref{fig:rate-scan-a} and \ref{fig:rate-scan-b})
to expose slow envelopes: over tens to hundreds of cycles the amide-bond
count drifts upward in a sequence of step-like increments separated by
partial regressions. These increments coincide with episodes in which the
hysteretic hydration controller keeps inflow off for extended periods,
allowing the ester pool to rebuild and then to be consumed in bursts of
aminolysis in subsequent low-water intervals. The regressions align with
longer wet intervals during which amide turnover transiently competes with
ester depletion. The resulting pattern exhibits imperfect phase locking
between hydration and chemistry (variable phase lag and amplitude),
intermittent plateaus in the amide fraction, and a sublinear long-term
trend indicative of a near-critical balance between amide hydrolysis and
ester-mediated amide formation under the chosen inflow/outflow schedule.

All simulation data (reaction network generated,
simulation trace with every event and their timestamps,
rule identifiers, the final state) are persistent, enabling
post-simulation analyses beyond time-series inspection.  In this study we
use these data to quantify frequencies of chemically interpretable
substructures (e.g., amide and ester motifs) and to track their evolution
across time. We do not perform trace compression or
counterfactual causal analysis~\cite{laurent-yang-fontana-counterfactual}
here. We note, however, that the content of the trace allows such techniques to be applied directly.
This include in particular
counterfactual, kinetics-aware causal reconstructions of minimal sufficient
event sets and directed influence relations in rule-based models~\cite{laurent-yang-fontana-counterfactual}.

\section{Conclusion} \label{sec: conclusion}

Stochastic simulation and rule-based modeling are two well
established methodologies, both frequently used in the
study of chemical reaction systems. The present work
introduces a stochastic simulation engine that combines these
approaches
on the basis of atomically explicit graph transformations.

At the core of the system is an implementation of Gillespie's direct
method that operates on explicit chemical graphs, thereby providing
atom-level resolution.  This resolution level is relevant in
particular for use cases where individual atoms must be tracked
through multiple reaction steps, such as the simulation of isotope
labeling experiments.  For this reason, the engine does not employ
techniques like tau-leaping~\cite{Gillespie:2001} for speeding up
simulations, since such approaches would compromise the ability to
track atoms explicitly.

The engine allows for creating detailed event traces which hold the
time-resolved information of rule applications and their embeddings
into the set of molecules in the system.  Additionally, a flexible
system of callback functions allows for the implementation of
elaborate experimental designs, including changes of rate constant
parameters and external fluxes in response to external control or
triggered by internal events.

The depsipeptide formation model presented serves as one illustrative example, demonstrating how generative chemical spaces can be coupled
directly to exact stochastic dynamics while retaining access to atom-level
structure, reaction embeddings, and dynamic environmental control. In such
systems, both the set of molecules and the reaction pathways emerge during
the simulation rather than being fixed in advance. 

The detailed event traces allowed
by the engine enable the inference of 
causal relationships between reactive events, generalizing work by the
Fontana lab on signaling system \cite{Cristescu_2019} using the Kappa
language \cite{boutillier2018kappa}.
This type of analysis reveals how feedbacks or macroscopic behavior such as
synchronized oscillations \cite{Forbes:2018} are implemented by the
microscopic events of the dynamic concurrent reactive system. It could
  be applied in future work, e.g., to guide the design of chemical reactive
systems (closed or open) that show predefined dynamic behavior
\cite{Harmsel:2023,LeCacheux:2025}.

In short, the core novelty of our contribution is to
enable stochastic simulation of generative chemical systems while retaining
atomically explicit structure and access to individual transformation
events, without requiring prior network construction. This shifts the focus
from predefined reaction networks to evolving spaces of chemically
admissible transformations, opening the possibility to systematically study
how such spaces are explored under stochastic dynamics.

The new simulation engine is built on top of the MØD computational
framework. MØD already provides an elaborate framework for partially
enumerating reaction networks, including an efficient caching
system. These features are extended here to seamlessly integrate with
the new stochastic simulation engine. Combined, this substantially
expands the scope of MØD as a modeling framework for rule-based
reaction systems.

We note that stochastic simulations of large, open-ended systems
are of interest also beyond chemistry for understanding
innovation and emergent features in complex systems. With its new
stochastic simulation engine, MØD provides an attractive option also
for simulations of general graph-based models.

\section{Declarations}

\subsection{Availability of Data and Materials}
The stochastic simulation engine is available as part of the latest
distribution of MØD from Github at
\url{https://github.com/jakobandersen/mod}, and the examples used in
this paper are available in a separate repository at
\url{https://github.com/jakobandersen/stochsim-examples}.

\subsection{Funding}
\label{sec: acknowledgement}

This work has received funding from the European Union's Horizon
Europe Doctoral Network programme under the Marie Skłodowska-Curie
grant agreement No.~101072930, from the Novo Nordisk Foundation grant
NNF19OC0057834, and from the Independent Research Fund Denmark,
Natural Sciences, grant DFF-0135-00420B.

\subsection{Competing Interests}
The authors declare that they have no competing interests.

\subsection{Author's Contributions}
EMHM: Methodology, Validation, Investigation, Data curation, Visualization, Writing - Review \& Editing.
JLA, RF, CF, DM, PFS: Methodology, Software, Validation, Writing - Review \& Editing, Supervision, Funding.

\bibliography{bibliography}

\clearpage
\newpage
\appendix

\section{ODE-Based Simulation of Degradation Dynamics of Atrazine in Soil}
\label{appendix:ODE}

For comparison with the stochastic simulations presented in the main text,
we additionally implemented a deterministic ODE model of the degradation
network shown in Fig.~\ref{fig:network}. Each molecular species
\texttt{A}--\texttt{P} corresponds to one node in the reaction graph, from
atrazine (\texttt{A}) to cyanuric acid (\texttt{P}). The model was
specified in Julia using the \texttt{DifferentialEquations.jl} and
\texttt{Catalyst.jl} packages, with the rate constants taken from
\cite{erickson1989degradation}. The full reaction network specification is
given in Listing~\ref{lst:atrazine_model}.

\begin{lstlisting}[caption={Reaction network for atrazine degradation in Julia using \texttt{Catalyst.jl}.}, label={lst:atrazine_model}]
rn = @reaction_network Atrazine begin
     @parameters k1 k2 k3
     (k2, k3, k1), A --> (B, C, D)
     (k3, k1, k1), B --> (E, F, H)
     (k2, k1, k1), C --> (E, G, I)
         (k2, k3), D --> (F, G)
         (k1, k1), E --> (J, K)
         (k3, k1), F --> (J, L)
         (k2, k1), G --> (J, M)
          (k2, k1), H --> (K, L)
         (k2, k1), I --> (K, M)
               k1, J --> O
         (k1, k1), K --> (N, O)
               k3, L --> O
               k2, M --> O
               k1, N --> P
               k1, O --> P
end
\end{lstlisting}

\begin{figure}
  \centering
  \includegraphics[width=0.65\linewidth]{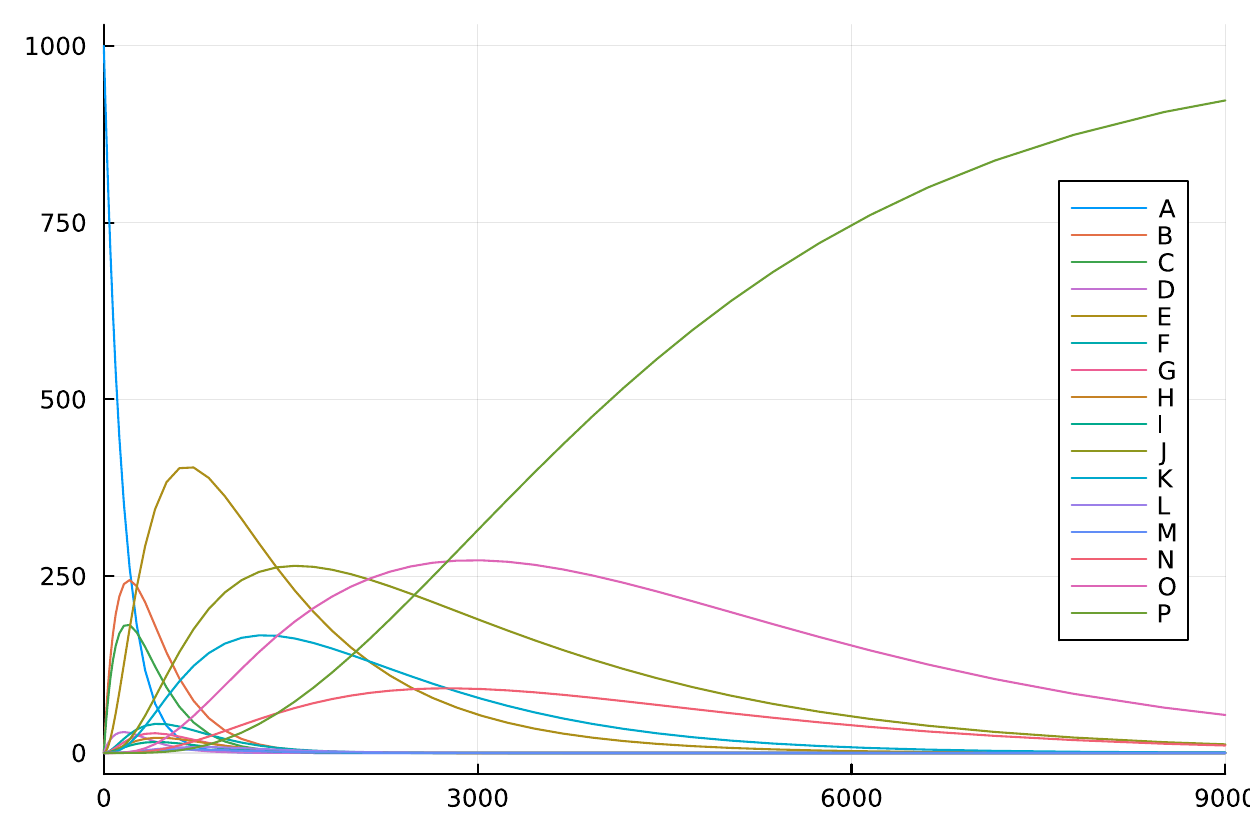}
  \caption{ODE-based degradation dynamics of atrazine in a closed system.
    Species \texttt{A}--\texttt{P} correspond to the intermediates in the
    reaction network of Fig.~\ref{fig:network}, with \texttt{A} = atrazine
    and \texttt{P} = cyanuric acid.  The system evolves monotonically
    toward \texttt{P}, in agreement with the closed stochastic simulation in Fig.~\ref{fig:atrazine-sim-closed}.}
  \label{fig:det}
\end{figure}

\begin{figure}
  \centering
  \includegraphics[width=0.7\textwidth]{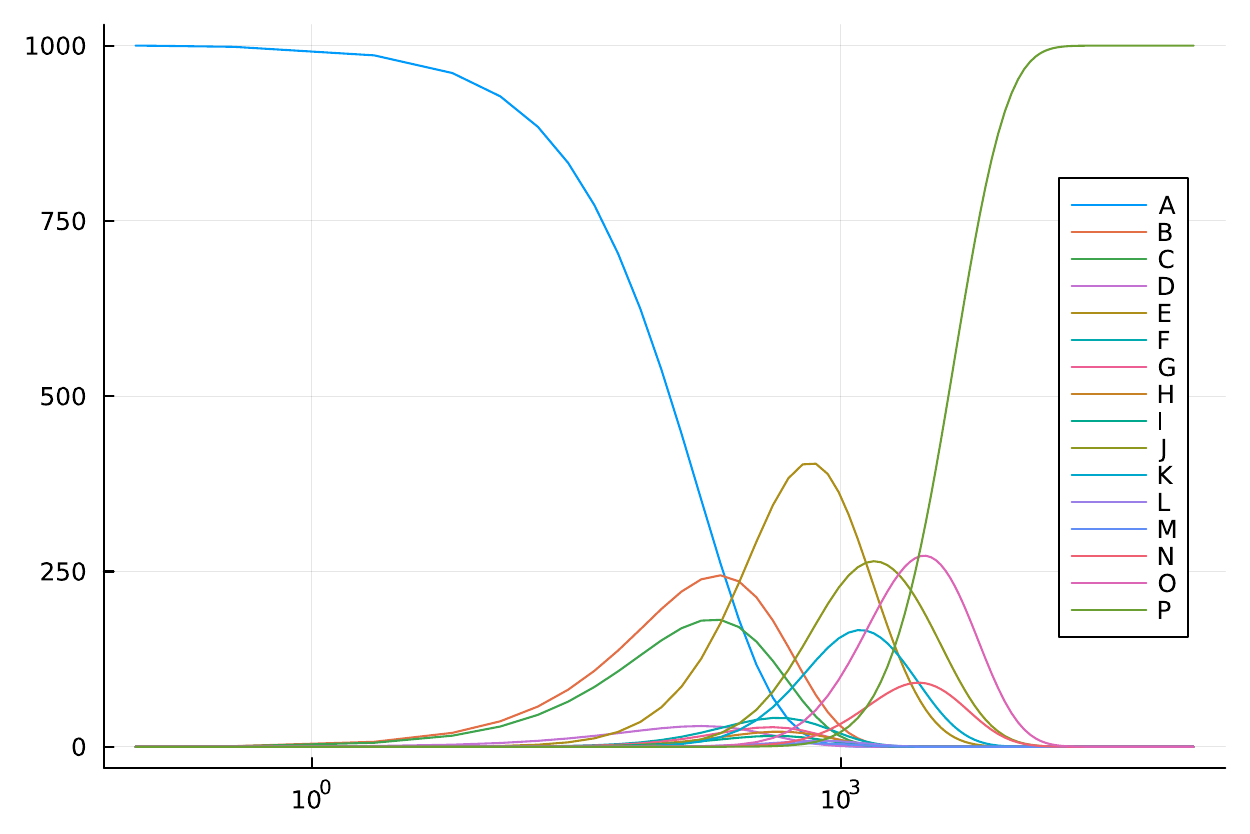}
  \caption{The same plot as in Fig.~\ref{fig:det}, but now with logarithmic time axis.}
  \label{fig:atrazine-dpo-closed-log}
\end{figure}

\begin{figure}
  \centering
  \includegraphics[width=0.7\textwidth]{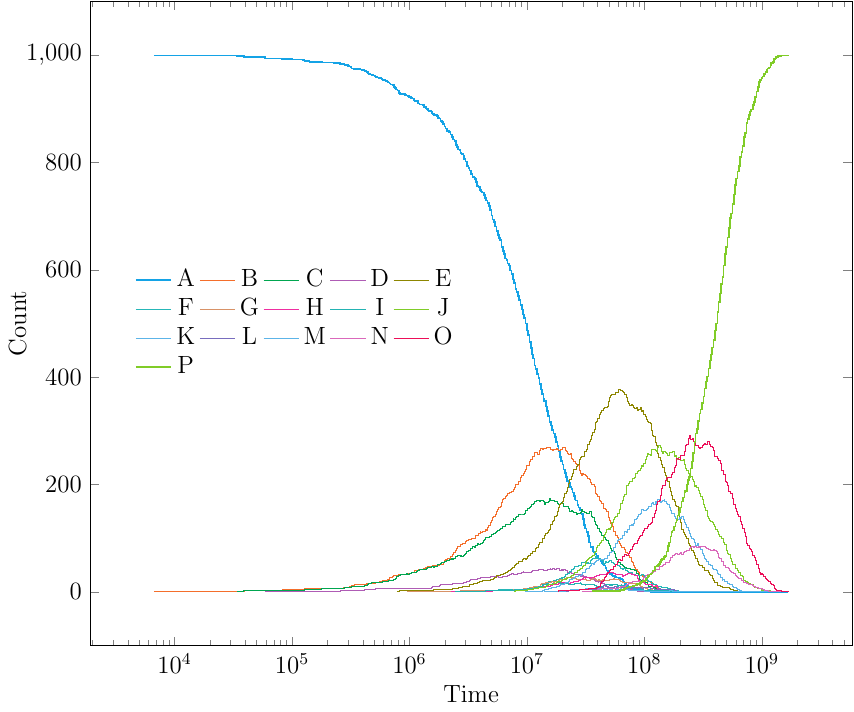}
  \caption{The same plot as in Fig.~\ref{fig:atrazine-sim-closed}, but now with logarithmic time axis.
   The starting conditions, $\texttt{A}=1000$, $\text{\texttt{B}--\texttt{P}}=0$,
   are not directly visible due to this axis scaling. 
 }
  \label{fig:atrazine-sim-closed-log}
\end{figure}

Fig.~\ref{fig:det} shows the time course of the deterministic ODE solution
obtained for an initial concentration of 1000 units of atrazine (\texttt{A}) and zero for all other species, simulated until $t=9000$.
As expected for a closed system, mass flows monotonically toward the stable product cyanuric acid
(\texttt{P}).  This deterministic result matches well the stochastic simulation of Fig.~\ref{fig:atrazine-sim-closed}, thereby corroborating the fidelity of our framework.
The similarity of plots becomes even clearer if we change the time axis from linear to logarithmic, as done in Fig.~\ref{fig:atrazine-dpo-closed-log} (ODE-based) and Fig.~\ref{fig:atrazine-sim-closed-log} (stochastic simulation), since most of the transients of the system then appear in the middle of the plots.


\clearpage
\section{Reaction Rules for Depsipeptide Formation}
\label{appendix:rules}
The rules used in Sec.~\ref{sec:stochastic-sim-depsi} are depicted in Fig.~\ref{fig:depsipep_rules}.
\begin{figure}[H]
  \centering
  \input{figures/rulesDepsipep/inc.tex}
  \caption{Graph transformation rules for the formation of depsipeptides.
    Each rule replaces a specific substructure in the left–hand side with a
    chemically valid product pattern in the right–hand side, preserving the
    atom labels and mass balance.}
  \label{fig:depsipep_rules}
\end{figure}

\end{document}